\documentclass[sigconf]{acmart}
\usepackage{commath}
\usepackage{multirow}
\usepackage[
  separate-uncertainty = true,
  multi-part-units = repeat
]{siunitx}
\usepackage{enumitem}

\AtBeginDocument{%
  \providecommand\BibTeX{{%
    \normalfont B\kern-0.5em{\scshape i\kern-0.25em b}\kern-0.8em\TeX}}}

\copyrightyear{2022} 
\acmYear{2022} 
\setcopyright{acmcopyright}\acmConference[LAK22]{LAK22: 12th International Learning Analytics and Knowledge Conference}{March 21--25, 2022}{Online, USA}
\acmBooktitle{LAK22: 12th International Learning Analytics and Knowledge Conference (LAK22), March 21--25, 2022, Online, USA}
\acmPrice{15.00}
\acmDOI{10.1145/3506860.3506883}
\acmISBN{978-1-4503-9573-1/22/03}

\begin{document}

\title{Understanding Distributed Tutorship in\\ Online Language Tutoring
}
\author{Meng Xia}
\affiliation{%
  \institution{School of Computing, KAIST}
  \country{South Korea}
}
\email{iris.xia@connect.ust.hk}

\author{Yankun Zhao}
\affiliation{%
  \institution{The Hong Kong University of Science and Technology}
  \country{Hong Kong}
}
\email{yzhaock@connect.ust.hk}

\author{Mehmet Hamza Erol}
\affiliation{%
 \institution{School of Computing, KAIST}
 \country{South Korea}
 }
 \email{mhamzaerol@kaist.ac.kr}
 
 \author{Jihyeong Hong}
\affiliation{%
  \institution{School of Computing, KAIST}
  \country{South Korea}
}
\email{z.hyeong@kaist.ac.kr}



\author{Juho Kim}
\affiliation{%
 \institution{School of Computing, KAIST}
 \country{South Korea}
 }
 \email{juhokim@kaist.ac.kr}
\renewcommand{\shortauthors}{Xia, et al.}

\newcommand{\para}[1]{\vspace{1mm}\noindent\textbf{#1}}
\newcommand{\numLearner}{15,959 }
\newcommand{\numSession}{278,191 }

\begin{abstract}
With the rise of the gig economy, online language tutoring platforms are becoming increasingly popular. They provide temporary and flexible jobs for native speakers as tutors and allow language learners to have one-on-one speaking practices on demand. However, the lack of stable relationships hinders tutors and learners from building long-term trust. ``Distributed tutorship''---temporally discontinuous learning experience with different tutors---has been underexplored yet has many implications for modern learning platforms. In this paper, we analyzed tutorship sequences of 15,959 learners and found that around 40\% of learners change to new tutors every session; 44\% learners change to new tutors while reverting to previous tutors sometimes; only 16\% learners change to new tutors and then fix on one tutor. We also found suggestive evidence that higher distributedness---higher diversity and lower continuity in tutorship---is correlated to slower improvements in speaking performance scores with a similar number of sessions. We further surveyed 519 and interviewed 40 learners and found that more learners preferred fixed tutorship while some do not have it due to various reasons. Finally, we conducted semi-structured interviews with three tutors and one product manager to discuss the implications for improving the continuity in learning under distributed tutorship.
\end{abstract}






\maketitle

\section{Introduction}
As the gig economy gains popularity~\cite{wright2017beyond, alkhatib2018laying}, where temporary, flexible jobs are commonplace for efficient resource allocation, new modes of teaching and learning spring up accordingly.
In particular, online language tutoring platforms (e.g., Ringle\footnote{\url{https://www.ringleplus.com/}}, Cambly\footnote{\url{https://www.cambly.com/english?lang=en}}, Preply\footnote{\url{https://preply.com/}}, and italki\footnote{\url{https://www.italki.com/}}) are becoming increasingly popular. These platforms provide temporary jobs for native speakers to work as part-time tutors, and at the same time, enable language learners to have one-on-one speaking sessions with native speakers anytime and anywhere~\cite{yeh2019speaking, kozar2014exploratory}.

Although it is flexible and convenient for both workers and consumers, the lack of a stable relationship and the long-term trust between workers and consumers is one of the concerns in the gig economy~\cite{hou2019measuring, zheng2020characteristics}. In particular, different from conventional learning with fixed instructors, online tutoring platforms allow learners to select tutors for every learning session. We define the learning experience in which learners distribute their learning time with different tutors as ``distributed tutorship'', implying learning discontinuously in time with different tutors. Since continuity and consistency are important dimensions of learning~\cite{galishnikova2015continuity}, and this mode of learning is unconventional in those dimensions, we want to investigate how distributed tutorship affects the learning process and outcome.

Previous work has explored the characteristics of different stakeholders in online language tutoring platforms, for example, the learners' demographics~\cite{kozar2014exploratory} and motivations~\cite{yung2020secondary} for learning, and tutors' conceptions and reasons for providing online tutoring services~\cite{werbinska2019english, yung2020most}. However, the influence of distributed tutorship on language learning is weakly investigated. In this paper, we investigate the following research questions:
\begin{itemize}[leftmargin=*]
    \item RQ1: What are the different patterns of distributed tutorship and their relationships with language learning improvements?
    \item RQ2: What are the reasons for learners to have different distributed tutorship settings?
    \item RQ3: What are the implications for educators, tutors, and platform designers to improve learners' language learning under distributed tutorship?
\end{itemize}

We explore the answers based on the context of a popular online English tutoring platform, Ringle\footnote{\url{https://www.ringleplus.com/en/student/landing/home}}, where English learners can select tutors from a list of native English speakers and have one-on-one speaking sessions on demand. For RQ1, we analyzed tutorship sequences (i.e., tutors across sessions) and learning performance data (i.e., scores given by tutors on grammar, vocabulary, fluency, and pronunciation for each learning session) of \numLearner learners. Inspired by previous research on categorizing learning sequences (e.g., ~\cite{kizilcec2013deconstructing}), we clustered learners' tutorship sequences and identified three patterns: 1) diverse tutorship: always changing to new tutors, 2) mixed tutorship: changing to new tutors while sometimes reverting to previous tutors, and 3) fixed tutorship: changing to new tutors and then fixing on one tutor. In particular, around 40\% of learners have the diverse tutorship, 44\% of learners have the mixed tutorship, and only 16\% of learners have the fixed tutorship. Learners who have more than 20 sessions all have the diverse tutorship. For learners who have \texttt{[2,20]} sessions, we further compared the learning performance of these three clusters and found that learners with fixed tutorship show higher learning gains than those with diverse tutorship and mixed tutorship. For learners who have more than 20 sessions, we propose ``distributedness'' to depict the degree of distributed tutorship by taking both diversity and continuity of the tutorship into consideration~\cite{goulet2017measuring}, calculated by the average Shannon entropy~\cite{shannon2001mathematical} of all the contiguous subsequences of a tutorship sequence. We found suggestive evidence showing that there is a negative correlation between tutorship distributedness and learning gains.

For RQ2, we surveyed 519 learners and interviewed 40 of them to understand the rationales behind having fixed or different tutors under distributed tutorship settings. Learners' main reason for wanting fixed tutors is the satisfaction with the current tutors and the continuity. For example, a fixed tutor can notice one's improvements or recurring language issues and give personalized instructions based on an in-depth understanding of the learner's previous learning experience. On the other hand, learners reported that the reason for having different tutors is they have not found their preferred tutors yet, and the schedule conflicts with their preferred tutors.

For RQ3, we conducted semi-structured interviews with three tutors and one product manager to understand how tutors and platforms could improve the learning experience under distributed tutorship. They explained the high degree of distributed tutorship to be primarily because most tutors are working part-time. They also provided potential considerations to improve the learning experience, especially improving the continuity of tutorship by encouraging tutors to give more standardized feedback, share learners' essential learning data among tutors, and use computer-aided tools to facilitate tutoring and learning.

The main contribution of this paper is the investigation of the emerging and trending mode of learning---distributed tutorship---in online language tutoring settings. We took a mixed-methods approach to identify major patterns of distributed tutorship and their relationship with language learning improvements, the reasons learners select different tutorship settings, and the implications for improving the learning experience under distributed tutorship. 
\section{Related Work}
In this section, we review previous work on online language tutoring platforms, continuity in language education, and learning sequence analysis.

\begin{table*}[]
\caption{The number of learners, mean, median, and SD of the number of tutors and the session tutor ratio in each range of sessions.}
\label{tab:learners}
\resizebox{\linewidth}{!}{
\begin{tabular}{l|lllllllllll}
\hline
Session ranges                                                           & \texttt{[2,10]}                                                                 & \texttt{[11,20]}                                                                & \texttt{[21,30]}                                                               & \texttt{[31,40]}                                                               & \texttt{[41,50]}                                                               & \texttt{[51,60]}                                                               & \texttt{[61,70]}                                                               & \texttt{[71,80]}                                                               & \texttt{[81,90]}                                                               & \texttt{[91,100]}                                                              & \texttt{[101,499]}                                                             \\ \hline
\# of learners                                                           & \multicolumn{1}{c}{9213}                                                   & \multicolumn{1}{c}{2950}                                                   & \multicolumn{1}{c}{1429}                                                  & \multicolumn{1}{c}{812}                                                   & \multicolumn{1}{c}{493}                                                                       & \multicolumn{1}{c}{290}                                                   & \multicolumn{1}{c}{200}                                                   & \multicolumn{1}{c}{147}                                                   & \multicolumn{1}{c}{107}                                                   & \multicolumn{1}{c}{65}                                                    & \multicolumn{1}{c}{253}                                                   \\ \hline
\begin{tabular}[c]{@{}l@{}}Mean, Median, SD \\ \# of tutors\end{tabular}       & \multicolumn{1}{c}{\begin{tabular}[c]{@{}c@{}}3.90,\\3,\\ 2.08\end{tabular}}                      & \multicolumn{1}{c}{\begin{tabular}[c]{@{}c@{}}10.93,\\11,\\ 3.91\end{tabular}}                     & \multicolumn{1}{c}{\begin{tabular}[c]{@{}c@{}}16.84,\\17,\\ 6.02\end{tabular}}                    & \multicolumn{1}{c}{\begin{tabular}[c]{@{}c@{}}22.48,\\24,\\ 8.77\end{tabular}}                    & \multicolumn{1}{c}{\begin{tabular}[c]{@{}c@{}}27.91,\\29,\\ 11.09\end{tabular}}                   & \multicolumn{1}{c}{\begin{tabular}[c]{@{}c@{}}32.79,\\33,\\ 13.51\end{tabular}}                   & \multicolumn{1}{c}{\begin{tabular}[c]{@{}c@{}}36.66,\\36,\\ 14.47\end{tabular}}                   & \multicolumn{1}{c}{\begin{tabular}[c]{@{}c@{}}43.45,\\44,\\ 17.68\end{tabular}}                   & \multicolumn{1}{c}{\begin{tabular}[c]{@{}c@{}}46.95,\\45,\\ 21.76\end{tabular}}                   & \multicolumn{1}{c}{\begin{tabular}[c]{@{}c@{}}49.72,\\48,\\ 22.90\end{tabular}}                   & \multicolumn{1}{c}{\begin{tabular}[c]{@{}c@{}}77.28,\\66,\\ 56.19\end{tabular}}                   \\ \hline
\begin{tabular}[c]{@{}l@{}}Mean, Median, SD\\ session tutor ratio\end{tabular} & \multicolumn{1}{c}{\begin{tabular}[c]{@{}c@{}}1.26,\\1.00,\\ 0.64\end{tabular}} & \multicolumn{1}{c}{\begin{tabular}[c]{@{}c@{}}1.67,\\1.25,\\ 1.39\end{tabular}} & \multicolumn{1}{c}{\begin{tabular}[c]{@{}c@{}}1.86,\\1.39,\\ 1.67\end{tabular}} & \multicolumn{1}{c}{\begin{tabular}[c]{@{}c@{}}2.14,\\1.48,\\ 2.29\end{tabular}} & \multicolumn{1}{c}{\begin{tabular}[c]{@{}c@{}}2.28,\\1.56,\\ 3.07\end{tabular}} & \multicolumn{1}{c}{\begin{tabular}[c]{@{}c@{}}2.30,\\1.65,\\ 2.23\end{tabular}} & \multicolumn{1}{c}{\begin{tabular}[c]{@{}c@{}}2.57,\\1.75,\\ 4.78\end{tabular}} & \multicolumn{1}{c}{\begin{tabular}[c]{@{}c@{}}2.40,\\1.77,\\ 2.50\end{tabular}} & \multicolumn{1}{c}{\begin{tabular}[c]{@{}c@{}}2.61,\\1.89,\\ 2.92\end{tabular}} & \multicolumn{1}{c}{\begin{tabular}[c]{@{}c@{}}2.80,\\1.90,\\ 2.76\end{tabular}} & \multicolumn{1}{c}{\begin{tabular}[c]{@{}c@{}}2.95,\\2.07,\\ 2.91\end{tabular}} \\ \hline
\end{tabular}
}
\end{table*}

\subsection{Online Language Tutoring Platforms}
With the rise of the gig economy~\cite{wright2017beyond, alkhatib2018laying}, online language tutoring platforms are becoming increasingly popular. On the one hand, they provide temporary and flexible jobs for native speakers to work as tutors. On the other hand, they allow language learners to have one-on-one speaking practices with native speakers at convenient times and locations~\cite{kozar2014exploratory}. Although it provides more flexibility and convenience for workers in the gig economy, the lack of stable relationships makes it hard for workers and consumers to build long-term trust, customary practice, and familiarity~\cite{hou2019measuring, zheng2020characteristics}. The impact and effectiveness of online language tutoring platforms in learning have rarely been explored yet.

Previous studies have investigated basic characteristics of online tutoring platforms or their stake. For example, researchers studied learners' motivations and goals~\cite{kozar2014exploratory, yung2020secondary} on online tutoring platforms, and they found that most of the learners on these platforms are adults, wanting to improve English for a better working experience or English certificate examinations. Other researchers found the reasons for tutors to work on these platforms include making money, helping others, and developing their professional skills. In addition, some work studied the characteristics of the websites of online tutoring platforms and revealed considerable thematic, structural and rhetorical similarities between the websites and a high presence of neoliberal ideology~\cite{kozar2015discursive}. Further work needs to be conducted to evaluate the relationship between learners and tutors on the platform and how it affects the learning outcome in online language tutoring platforms.

\subsection{Continuity in Education}
Previous studies pointed out the importance of continuity~\cite{galishnikova2015continuity} in language education. Most of the studies investigated the continuity in curriculum design across different levels of education~\cite{kardaleska2009continuity, cepon2012higher}, e.g., how to maintain the curriculum continuity from primary to secondary school or from secondary school to higher education. Some investigated the content and methodology for English learning within one level of education. For example, Galishnikova~\emph{et al.} investigated the importance of continuity in higher education~\cite{galishnikova2015continuity}. Their experiments showed that undergraduates in the experimental group (N=15) with continuous learning content and methodology achieve higher proficiency in English (e.g., writing abstracts) than the control group (N=22).

A few works discussed the continuity issues in other domains. Ostrow~\emph{et al.} found that interleaving cognitive content in online adaptive tutoring systems in math is beneficial for students with low skills~\cite{ostrow2015blocking}.
However, there is no work before systematically investigating the influence on learning brought by distributed tutorship, which involves both diversity and continuity in tutorship across learning sessions.


\subsection{Learning Sequence Analysis}
Considerable research efforts have been made to detect and classify learning sequences. One large branch of research identified different learner groups on MOOCs by clustering different learning sequences~\cite{kizilcec2013deconstructing, sinha2014capturing, fu2016visual, chen2016dropoutseer, chen2018viseq}. For example, Kizilcec~\emph{et al.}~\cite{kizilcec2013deconstructing} analyzed the disengagement patterns of MOOC learners by first encoding learners' weekly assignment completion behaviors into auditing, behind, and on track, then grouping the encoded weekly behavior sequences into four categories with different engagement patterns: completing, auditing, disengaging, and sampling. DropoutSeer~\cite{chen2016dropoutseer} predicted learners' dropout behaviors on MOOCs by clustering learners' video watching behaviors. ViSeq~\cite{chen2018viseq} analyzed various learning behaviors, including video watching, forum discussion, and assignment submitting, and found that learners with different learning outcomes exhibit different learning patterns before the final exam. Most of the learning sequences of MOOC data analyzed in this branch can be aligned to the same length based on the fixed curriculum on MOOCs. However, tutorship sequences in online platforms are of different lengths as no fixed curriculum is provided or required, bringing new challenges to the analysis.

Another branch of research involved learning sequences analysis on problem-solving behaviors. For example, Shanabrook~\emph{et al.} proposed using motifs to identify students' frequent patterns and engagements states while working toward problems~\cite{shanabrook2010identifying}. Xia~\emph{et al.} analyzed how students solve a series of questions in programming exercises and used the Markov Chain to model and recommend the learning paths based on learners' submission outcomes (i.e., pass or not)  ~\cite{xia2019peerlens}. They further analyzed learners' detailed problem-solving steps in math problems~\cite{xia2020qlens}. However, these approaches focus more on local features (e.g., transitions between learning states) instead of features of the whole learning sequence. Inspired by the previous sequence modeling methods, we propose a new encoding scheme to encode the tutorship sequence and a new measurement of distributedness of tutorship to analyze tutorship sequences of various lengths while taking global features (e.g., tutorship diversity and continuity) of each sequence into consideration.

\section{Distributed Tutorship Patterns and their Relationship with Learning Improvements}
\begin{figure*}
\centering
  \includegraphics[width=\linewidth]{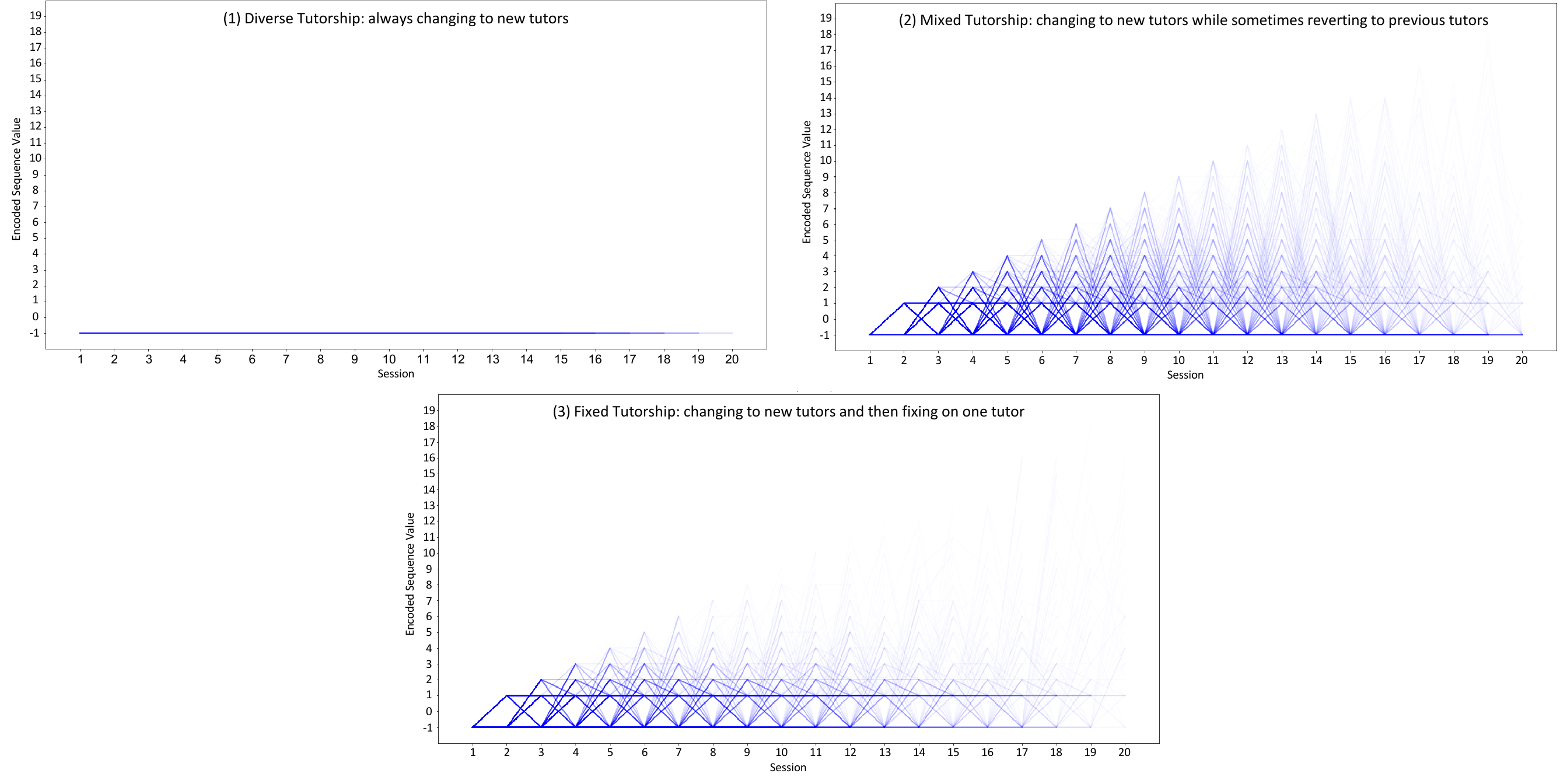}
  \caption{Three clusters of different distributed tutorship patterns for learners have \texttt{[2,20]} sessions. The x-axis represents the session number, and the y-axis represents the encoded values in the sequence (i.e., -1 means the current tutor is new; 1 means the current tutor appeared one session prior; 2 means the current tutor appeared two sessions prior, and so on). Each line demonstrates one learner's tutorship sequence. Subfigure (1) shows the diverse tutorship where learners always change to new tutors ($y = -1$ across sessions); (2) shows the mixed tutorship, where learners change to new tutors and sometimes revert to previous tutor ($y = -1$ or other values), and (3) shows fixed tutorship, where learners first try different tutors and then fix on one tutor ($y = -1$ is darker at the beginning while $y = 1$ later on is more salient than $y = -1$ and other lines)}
  \label{fig:clusters}
\end{figure*}

\label{sec:data}
This section investigates RQ1: What are the different patterns of distributed tutorship and their relationships with language learning improvements? We first introduce the online tutoring platform that our research builds on, then present learners' different distributed tutorship patterns. Finally, we analyze their relationships with learning improvements.

The data we analyzed comes from Ringle, an online English tutoring platform. On Ringle, learners can choose a tutor and class time for 1:1 online speaking sessions with native English speakers. Learners on this platform are primarily adults who have the intermediate-to-advanced level of English. In particular, we analyzed the learners' tutorship sequences (i.e., the sequence of tutors across sessions) and learning improvements (i.e., the slopes of tutors' scores on grammar, vocabulary, fluency, and pronunciation across sessions) of \numLearner learners in Ringle who had at least two sessions with tutor scores.

\subsection{Highly Active Distributed Tutorship on the Platform}

Our data demonstrate that a high degree of distributed tutorship is taking place on the platform. In total, our data contain \numSession sessions, and the average number of sessions per learner is 17 (min:2, max: 499). The distribution of the number of learners, the mean, median, and SD of the number of tutors and the session tutor ratio in each session range is listed in Table~\ref{tab:learners}. The session tutor ratio is calculated by \( \frac{\textrm{\# of sessions}}{\textrm{\# of unique tutors}} \) for each learner. For example, in the first column, for the learners who have \texttt{[2,10]} sessions, they have 3.9 tutors on average, and each tutor taught 1.26 sessions on average. From the table, we can see that for learners in each range, the mean session tutor ratio is less than 3, which indicates the same tutor taught a learner less than three sessions on average.

\subsection{Tutorship Sequence Clustering}
\begin{table*}[]
\caption{Statistics of the three clusters (diverse tutorship, mixed tutorship, fixed tutorship) of learners who had \texttt{[2,20]} sessions. *Mixed tutorship is the cluster for learners who had more than 20 sessions.}
\label{tab:cluster_info}
\resizebox{\linewidth}{!}{
\begin{tabular}{l|lccc}
\hline
Session ranges              & Cluster names     & \multicolumn{1}{l}{\# of learners (percentage)} & \multicolumn{1}{l}{Mean (SD) \# of sessions} & \multicolumn{1}{l}{Mean (SD) \# of tutors} \\ \hline
\multirow{3}{*}{\texttt{[2,20]}} & Diverse tutorship & 6437 (40\%)                                     & 4.64 (3.38)                                                  & 4.64 (3.38)                                                \\ \cline{2-5} 
                            & Mixed tutorship   & 3169 (20\%)                                     & 11.36 (4.97)                                                 & 8.56 (4.20)                                                \\ \cline{2-5} 
                            & Fixed tutorship   & 2557 (16\%)                                     & 7.98 (5.11)                                                  & 4.37 (3.33)                                                \\ \hline
\texttt{[21,499]}                & *Mixed tutorship  & 3796 (24\%)                                     & 47.54 (37.14)                                                & 28.22 (23.73)                                              \\ \hline
\end{tabular}}
\end{table*}
Inspired by previous research on clustering learning sequences (e.g., \cite{kizilcec2013deconstructing}), we categorized learners' tutorship sequences for different distributed tutorship patterns. Each learner has a tutorship sequence, e.g., $\langle$\textit{abcbc}$\rangle$, where \textit{a, b, c} are three distinct tutors. We encoded the tutorship sequence according to two rules: (1) encoding a tutor using a positive number to show the session offset from the last appearance of this tutor (i.e., 1 means the current tutor appeared one session prior; 2 means two sessions prior, and so on); (2) if the tutor appears at the first time, encoding the tutor with a negative number (e.g., -1 in our paper) to distinguish it from positive session offsets to set up a larger euclidean distance with the encoding of existing tutors. For example, $\langle\textit{aaaaaa}\rangle$ -> $\langle\textit{(-1)11111}\rangle$; $\langle\textit{aaabbb}\rangle$ -> $\langle\textit{(-1)11(-1)11}\rangle$; $\langle\textit{ababab}\rangle$ -> $\langle\textit{(-1)(-1)2222}\rangle$; $\langle\textit{abcabc}\rangle$ -> $\langle\textit{(-1)(-1)(-1)333}\rangle$; $\langle\textit{abcdef}\rangle$ -> $\langle\textit{(-1)(-1)(-1)(-1)(-1)(-1)}\rangle$. The reasons for this representation are three-fold. First, it encodes both the diversity (i.e., changing new tutors) and continuity (i.e., choosing the same tutor in consecutive sessions) in tutorship. Second, with this numerical representation, the distance between two sequences can be computed. Third, the tutorship sequences of a group of learners can be further visualized for interpretation. These reasons are explained in detail in the following paragraphs.

With the numerical representations of tutorship sequences, we then applied the dynamic time warping (DTW) algorithm~\cite{sakoe1978dynamic} to calculate the similarity between each pair of tutorship sequences. We chose this algorithm since it can calculate the distance between sequences of different lengths.
Consider a sequence $X$ with length $m$ and a sequence $Y$ with length $n$. The idea of this algorithm is to find a path in an $m$-by-$n$ matrix, with the distance between $X_{i\in m}$ and $Y_{j\in n}$ as value, that has the minimum cumulative distance~\cite{berndt1994using}. In our case, we use Euclidean distance as the distance function.
We further used k-medoids clustering~\cite{kaufman1990partitioning} to cluster the sequences based on their similarity scores.
We applied k-medoids clustering in different subgroups of learners based on their number of sessions (e.g., \texttt{[2,20]}, \texttt{[21,40]}, \texttt{[2,30]}, \texttt{[31,60]}) instead of applying to all the sequences in \texttt{[2,499]}. The reason is that the DTW algorithm cannot capture the difference between tutorship sequences accurately when their lengths differ significantly and predominantly affect the distance calculation. 
In k-medoids, the k (i.e., number of clusters) should be specified, and we tested each k from 1 to 20. 
For all the trials of different ranges and k values, we ran the Silhouette cluster validation~\cite{rousseeuw1987silhouettes} to test the ``goodness of fit'' of the clusters ~\cite{kizilcec2013deconstructing} and interpreted the clustering results using data visualization. Finally, we identified three clusters, that is, diverse tutorship, mixed tutorship, and fixed tutorship. The reasons for having the three clusters are two-fold. First, they have relatively good silhouette scores. The average silhouette score of the three clusters for learners who have \texttt{[2,20]} sessions is 0.54. The Silhouette score ranges from \texttt{[-1,1]}, and the higher the score, the better the clustering results. For learners who have more than 20 sessions, we clustered them into one cluster (i.e., mixture tutorship) since there is no clear pattern of other clusters. We use 20 as the dividing line as \texttt{[2,20]} comprises the majority (76\%) of the learners with relatively good Silhouette scores. Second, they present meaningful distributed tutorship patterns for learners, explained in the next paragraph. The features of each cluster, including the number of learners, sessions, and tutors, are listed in Table~\ref{tab:cluster_info}. 


We further visualized the tutorship sequence of each of the three clusters for learners who have \texttt{[2,20]} sessions in Figure~\ref{fig:clusters}. In Figure~\ref{fig:clusters}, the x-axis represents the session number, and the y-axis represents the encoded values in the sequence (i.e., -1 means the current tutor is new; 1 means the current tutor appeared one session prior; 2 means the current tutor appeared two sessions prior, and so on). Each line demonstrates one learner's tutorship sequence, and each subfigure (1), (2), and (3) shows the summary of the sequences of learners in that cluster. We explain the three clusters as follows.

\begin{itemize}[leftmargin=*]
    \item \noindent Diverse Tutorship: always changing to new tutors. There are 40\% of learners in this cluster, and learners in this cluster always change to new tutors without seeking to learn with previous tutors. As shown in Figure~\ref{fig:clusters}(1), there is a straight line of $y = -1$, which means that each tutor is distinct within the sequence for each learner in this cluster.
    \item \noindent Mixed Tutorship: changing to new tutors while sometimes reverting to previous tutors. There are 20\% learners who have \texttt{[2,20]} sessions in this cluster. Learners in this cluster change to new tutors and sometimes revert to tutors they had several sessions prior. From Figure~\ref{fig:clusters}(2), we can see that this group has a thick line in $y = -1$, which means they change to new tutors. Meanwhile, there are also some peaks with positive values along the x-axis, which means they revert to previous tutors from time to time.
    \item \noindent Fixed Tutorship: first try different tutors and then fixing on one tutor. Learners from this cluster first change to new tutors as the line of $y = -1$ is darker at the beginning. They then fix on one tutor, as seen in Figure~\ref{fig:clusters}(3) that $y = 1$ further on is more salient than $y = -1$ and other lines.
\end{itemize}

\subsection{Relationship with Learning Improvements}
We now report the relationship between different distributed tutorship patterns and learning improvements. This analysis aims to reveal if certain types of distributed tutorship patterns are associated with higher learning gains. Due to the lack of control and various possible confounding factors, we do not claim causality between the distributed tutorship pattern and the learning gain. But we believe the correlation analysis can still provide meaningful insights into the design of distributed tutorship experience. We collected tutors' scores on fluency, grammar, vocabulary, and pronunciation for each learning session on Ringle. Learning improvements are represented using the slopes of the regression lines of tutors' scores on grammar, vocabulary, fluency, and pronunciation across learning sessions. 

\begin{figure*}
\centering
  \includegraphics[width=\linewidth]{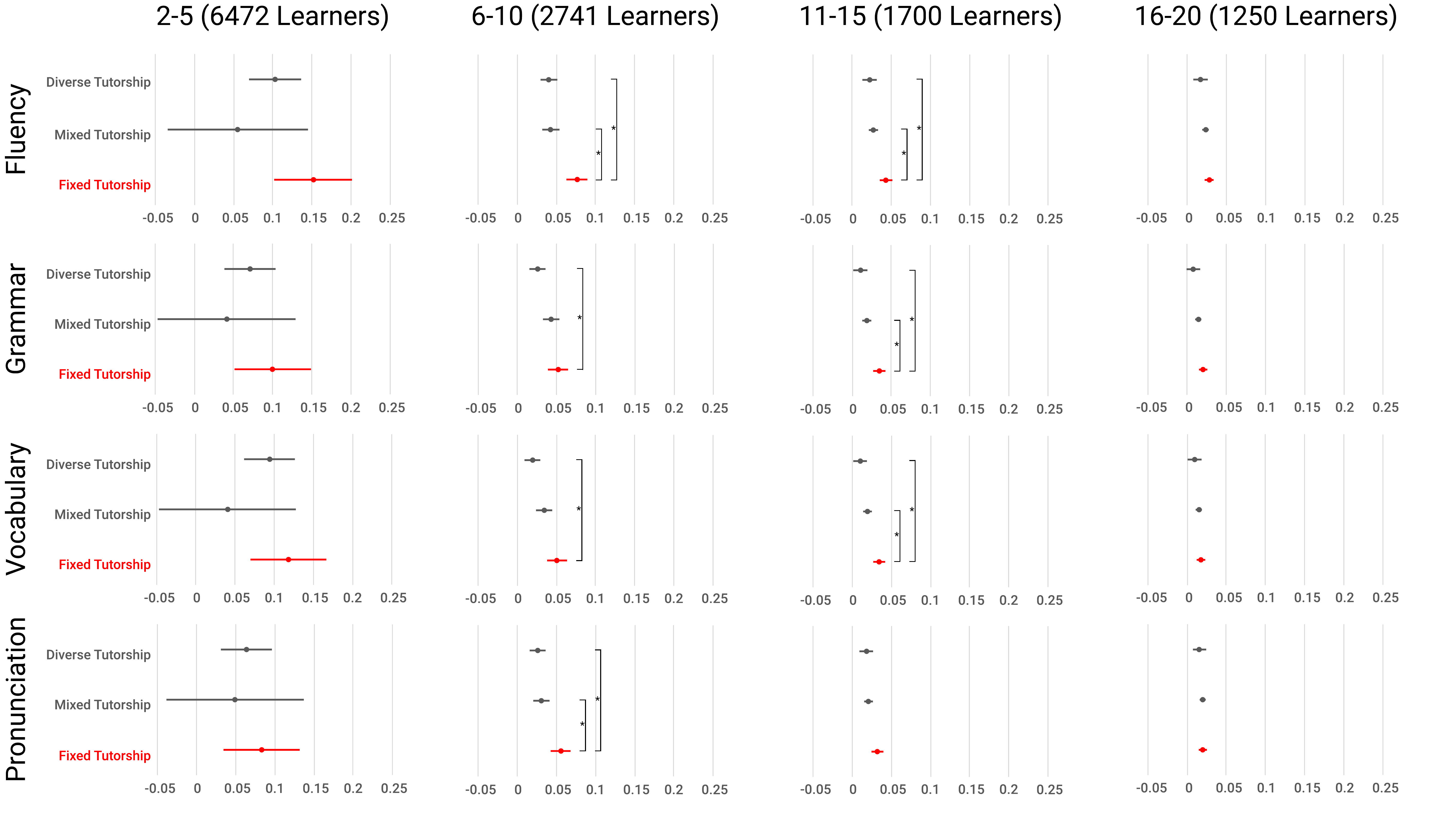}
  \caption{Mean with 95\% C.I. of the slopes of the regression lines of fluency scores, grammar scores, vocabulary scores, and pronunciation scores of learners in three clusters (i.e., diverse tutorship, mixed tutorship, and fixed tutorship) with session ranges \texttt{[2,5] (6472)}, \texttt{[6,10]} (2741), \texttt{[11,15]} (1700), and \texttt{[16,20]} (1250). $*:p<.05 $.} 
  \label{fig:comparison}
\end{figure*}

\para{Learners who have \texttt{[2,20]} sessions.} For learners who have \texttt{[2,20]} sessions, we compared learning improvements of the three clusters. Since the number of sessions would affect the learning outcome (i.e., more sessions generally mean the learner studied more and longer on the platform, possibly with a sustained level of motivation), we further divide the learners into fine-grained ranges based on their number of sessions: \texttt{[2,5]}, \texttt{[6,10]}, \texttt{[11,15]}, \texttt{[16,20]}. We ran a one-way analysis of variance (ANOVA) on the learning improvements of learners in each cluster. This is followed by Tukey Honest Significant Differences (HSD) test for post hoc pair-wise cluster comparison~\cite{hothorn2006handbook} and the results are listed in Figure~\ref{fig:comparison}. We found that learners from the fixed tutorship cluster have higher learning improvements in fluency, grammar, vocabulary, and pronunciation in all ranges. In particular, in the session ranges \texttt{[6,10]} and \texttt{[11,15]}, the mean value of the fixed tutorship cluster has a significant difference over the mean values of the mixed tutorship and diverse tutorship clusters. It reflects to a certain extent that learners who had fixed tutors result in higher learning improvements than learners who learned from different tutors time by time.

\begin{table*}[]
\caption{Spearman correlation coefficient and p-value for the relationship between distributedness and the slopes of the regression lines of fluency scores, grammar scores, vocabulary scores, and pronunciation scores of learners in each range, respectively. $*:p<.05 $.}
\label{tab:correlation}
\begin{tabular}{|l|c|c|c|c|c|}
\hline
Ranges         & \multicolumn{1}{l|}{\# of learners} & \multicolumn{1}{l|}{\begin{tabular}[c]{@{}l@{}}(Correlation, p) \\ of distributedness \\ \& fluency\_slope\end{tabular}} & \multicolumn{1}{l|}{\begin{tabular}[c]{@{}l@{}}(Correlation, p) \\ of distributedness \\ \& grammar\_slope\end{tabular}} & \multicolumn{1}{l|}{\begin{tabular}[c]{@{}l@{}}(Correlation, p)\\ of distributedness \\ \& vocabulary\_slope\end{tabular}} & \multicolumn{1}{l|}{\begin{tabular}[c]{@{}l@{}}(Correlation, p) of\\ distributedness \&\\ pronunciation\_slope\end{tabular}} \\ \hline
\texttt{[2,5]}      & 6472                                & (-0.0427, 0.0006)*                                                                                                       & (-0.0184, 0.1378)                                                                                                       & (-0.0341, 0.0060)*                                                                                                         & (-0.0316, 0.0109)*	                                                                                                             \\ \hline
\texttt{[6,10]}    & 2741                                & (-0.0463, 0.0154)*                                                                                                       & (-0.0293, 0.1246)                                                                                                      & (-0.0349, 0.0674)                                                                                                         & (-0.0190, 0.3201)                                                                                                             \\ \hline
\texttt{[11,15]}   & 1700                                & (-0.0623, 0.0102)*                                                                                                       & (-0.1153, 0.0000)*                                                                                                         & (-0.1014, 0.0000)*                                                                                                         & (-0.0473, 0.0510)                                                                                                             \\ \hline
\texttt{[16,20]}    & 1250                                & (-0.0926, 0.0010)*                                                                                                       & (-0.1168, 0.0000)*                                                                                                          & (-0.0863, 0.0023)*                                                                                                         & (-0.0560, 0.0477)*                                                                                                             \\ \hline
\texttt{[21,25]}   & 860                                 & (-0.0669, 0.0498)*                                                                                                        & (-0.0651, 0.0564)                                                                                                      & (-0.0615, 0.0715)                                                                                                          & (-0.0428, 0.2100)                                                                                                              \\ \hline
\texttt{[26,30]}   & 569                                 & (-0.0490, 0.2437)                                                                                                        & (-0.0513, 0.2218)                                                                                                       & (-0.0827, 0.0487)*                                                                                                          & (-0.0295, 0.4824)                                                                                                             \\ \hline
\texttt{[31,35]}   & 445                                 & (-0.0987, 0.0374)*                                                                                                       & (-0.0551, 0.2460)                                                                                                        & (-0.1339, 0.0047)*                                                                                                          & (-0.0032, 0.9469)                                                                                                              \\ \hline
\texttt{[36,40]}   & 367                                 & (-0.0617, 0.2380)                                                                                                         & (0.0341, 0.5148)                                                                                                        & (-0.0034, 0.9489)                                                                                                          & (0.0069, 0.8945)                                                                                                              \\ \hline
\texttt{[41,45]}   & 283                                 & (-0.1532, 0.0099)*                                                                                                       & (-0.1314, 0.0270)*                                                                                                      & (-0.0968, 0.1041)                                                                                                          & (-0.0999, 0.0933)                                                                                                             \\ \hline
\texttt{[46,50]}   & 210                                 & (-0.0751, 0.2784)                                                                                                        & (-0.1058, 0.1263)                                                                                                        & (-0.1462, 0.0342)*                                                                                                          & (-0.1529, 0.0267)*                                                                                                            \\ \hline
\texttt{[51,55]}   & 140                                 & (-0.0909, 0.2856)                                                                                                        & (-0.0504, 0.5541)                                                                                                       & (-0.1233, 0.1466)                                                                                                          & (0.0196, 0.8178)                                                                                                             \\ \hline
\texttt{[56,60]}   & 150                                 & (-0.1013, 0.2174)                                                                                                        & (-0.0259, 0.7532)                                                                                                       & (-0.1544, 0.0593)                                                                                                          & (0.0993, 0.2264)                                                                                                              \\ \hline
\texttt{[61,65]}   & 108                                 & (0.0361, 0.7107)                                                                                                        & (0.0402, 0.6799)                                                                                                        & (0.0609, 0.5310)                                                                                                            & (0.0094, 0.9229)                                                                                                              \\ \hline
\texttt{[66,70]}   & 92                                  & (-0.1674, 0.1107)                                                                                                        & (-0.1601, 0.1275)                                                                                                       & (-0.0890, 0.3990)                                                                                                          & (0.0324, 0.7588)                                                                                                              \\ \hline
\texttt{[71,75]}   & 66                                  & (-0.2339, 0.0587)                                                                                                        & (-0.1647, 0.1864)                                                                                                       & (-0.1527, 0.2211)                                                                                                          & (-0.0828, 0.5086)                                                                                                             \\ \hline
\texttt{[76,80]}   & 81                                  & (-0.1126, 0.3169)                                                                                                        & (-0.1233, 0.2727)                                                                                                       & (-0.0212, 0.8508)                                                                                                          & (-0.1581, 0.1587)                                                                                                             \\ \hline
\texttt{[81,85]}   & 57                                  & (-0.0204, 0.8802)                                                                                                        & (-0.0266, 0.8444)                                                                                                       & (-0.0956, 0.4793)                                                                                                          & (-0.1709, 0.2038)                                                                                                             \\ \hline
\texttt{[86,90]}   & 50                                  & (-0.2976, 0.0358)*                                                                                                       & (-0.1517, 0.2930)                                                                                                       & (-0.3222, 0.0225)*                                                                                                          & (-0.0769, 0.5954)                                                                                                             \\ \hline
\texttt{[91,95]}   & 37                                  & (0.0730, 0.6676)                                                                                                          & (0.0059, 0.9722)                                                                                                        & (-0.0266, 0.8760)                                                                                                           & (0.1221, 0.4716)                                                                                                              \\ \hline
\texttt{[96,100]}  & 28                                  & (-0.1450, 0.4615)                                                                                                         & (-0.1149, 0.5603)                                                                                                       & (-0.0285, 0.8857)                                                                                                          & (0.2228, 0.2545)                                                                                                              \\ \hline
\texttt{[101,499]} & 253                                 & (-0.3402, 0.0000)*                                                                                                          & (-0.2776, 0.0000)*                                                                                                         & (-0.2278, 0.0000)*                                                                                                            & (-0.2445, 0.0000)*                                                                                                               \\ \hline
\end{tabular}
\end{table*}


\begin{figure*}
\centering
  \includegraphics[width=\linewidth]{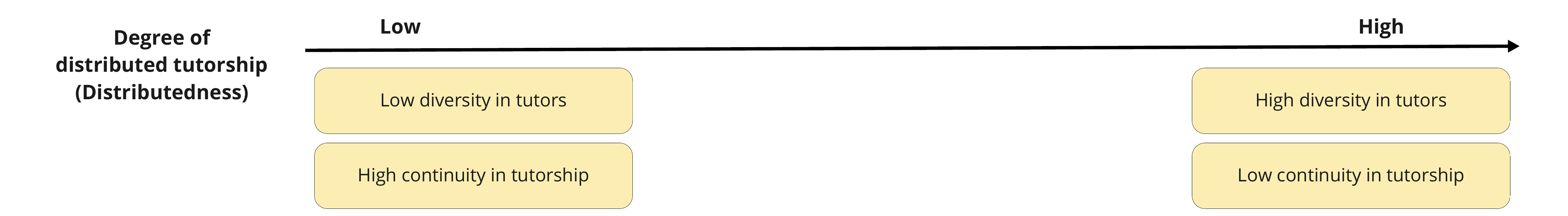}
  \caption{Two dimensions of distributed tutorship: diversity in tutors and continuity in tutorship.} 
  \label{fig:dimensions}
\end{figure*}


\para{Learners who have \texttt{[21,499]} sessions.} Since all learners who have more than 20 sessions are in one cluster, mixed tutorship, to further analyze the characteristics of their distributed tutorship and its relationship with learning improvements, we quantify the ``distributedness'' of each learner's tutorship sequence. ``Distributedness'' is proposed to capture the diversity and continuity in the tutorship. A low tutorship distributedness means a low diversity in tutors and a high continuity in tutorship, and a high tutorship distributedness represents a high diversity in tutors and a low continuity in tutorship, as shown in Figure~\ref{fig:dimensions}.
With a tutorship sequence (e.g.,  $\langle$\textit{abcbc}$\rangle$, where \textit{a, b, c} are three different tutors), we propose the average Shannon entropy of all the contiguous subsequences of one tutor sequence to represent the distributedness of the tutorship sequence. Given a tutorship sequence $X$, let $S = \{S_1, S_2, ..., S_n\}$ be a set of contiguous subsequences of the sequence $X$, and its distributedness is defined as:
$Distributedness(X)=\frac{1}{n}\sum_{S_i\in S}H(S_i)$. 
In the equation above, $H(S_i)$ is the Shannon entropy of subsequence $S_i$.
$H(S_i)=\sum_{s\in Uniq(S_i)}P(\frac{Count(s)}{|S_i|})\log P(\frac{Count(s)}{|S_i|})$,
where $Uniq(S_i)$ is the set of unique tutors in the subsequence; $Count(s)$ is the total number of unique items $s$; $|S_i|$ is the total number of items in the subsequence $S_i$.
For example, if a sequence is $\langle\textit{aabb}\rangle$, $Distributedness (\langle\textit{aabb}\rangle) = (H (\langle\textit{a}\rangle) + H (\langle\textit{a}\rangle) + H (\langle\textit{b}\rangle) + H (\langle\textit{b}\rangle) + H (\langle\textit{aa}\rangle) + H (\langle\textit{ab}\rangle) + H (\langle\textit{bb}\rangle) + H (\langle\textit{aab}\rangle) + H (\langle\textit{abb}\rangle) + H (\langle\textit{aabb}\rangle))/10$.
The justifications of the calculation are as follows. First, by surveying previous research, we found that entropy is commonly used to calculate diversity and regularity of learning activities ~\cite{allen2017d,hecking2015uncovering,poquet2017effective,jovanovic2020supporting,hsiao2015topic,saqr2020high}. Therefore, entropy itself can represent the diversity of the tutorship. Second, we use the average entropy of all the subsequences instead of the entropy of the whole sequence because we want to further take the continuity in tutorship into consideration. If all the subsequences of a tutorship sequence have low entropy values, then it means the learner did not change tutors frequently in all phases of learning and maintained continuous learning. For example, although $\langle\textit{aaabbb}\rangle$ and $\langle\textit{ababab}\rangle$ have the same entropy value, $\langle\textit{aaabbb}\rangle$ has a lower distributedness value than $\langle\textit{ababab}\rangle$ according to our metric, which can reflect the difference in the continuity in the tutorship between these two sequences. A similar idea was used to calculate the regularity in travel patterns~\cite{goulet2017measuring}. 

We then calculated the Spearman correlation coefficient of their learning improvements with the distributedness in different groups based on number of sessions: \texttt{[2,5]}, \texttt{[6,10]}, ... \texttt{[31,35]}, \texttt{[36,40]}, ..., \texttt{[91,95]}, \texttt{[96,100]}, \texttt{[101,499]}. The reason we conducted correlation analysis respectively within different session ranges is because the number of sessions affects learning improvement, thus comparing learners who had similar numbers of sessions helps to better control this variable and ensure a fair comparison. The last group is \texttt{[101,499]} because with the long-tail distribution there is limited data in every five sessions in that range. The results of the correlation value and p-value are listed in Table~\ref{tab:correlation}, where $p < 0.05$ represents there is a statically significant correlation. 

We can see that in all ranges, the correlation value of distributedness and the slopes of the regression lines of fluency scores, grammar scores, vocabulary scores, and pronunciation scores are almost all negative, though the absolute value is not large. In particular, in ranges \textit{[2,5], [6,10], [11, 15], [16, 20]}, there are significant negative correlations, which aligns with our previous comparisons on clusters and shows the effectiveness of the measure of ``distributedness''. What worth mentioning is that, for learners with more than 100 sessions, the correlation value is around or larger than 0.25 in all four aspects with $p < 0.05$, which means there is a weak negative correlation between tutorship distributedness and learning improvements.

\section{Reasons for Distributed Tutorship}
To further understand the reasons behind the results in the data analysis and different distributed tutorship settings (RQ2), we conducted a survey and interviews with learners from Ringle.

\subsection{Surveys and Interviews with Learners}
\label{sec:survey}
We posted survey questions on the Ringle's web page. The survey questions included their demographics (e.g., age, gender) and one open-ended question:
``What are the reasons for you to have fixed or different tutors?'' In total, we got responses from 519 learners (127 males, 381 females, 11 not disclosed, age: $\SI{31.81 \pm 7.25}{}$) on Ringle whose first language was not English. Their occupation distribution is: employed for wages (362), student (92), out of work (32), self-employed (25), homemaker (5), military (1), retired (1), or unable to work (1).
Their educational background distribution is bachelor's degree (298), master's degree (126), some college credits without degree (39), professional degree including doctor or lawyer (22), doctoral degree (20), graduated elementary/middle/high school (9), none (2), or vocational training (1). For 402 learners who have records on the platform, the number of sessions they took is $\SI{33.22 \pm 52.46 }{}$ (min 1, max 501). Among the 519 learners, 40 of them (12 males, 28 females, age: $\SI{33.98 \pm 7.01}{}$) who signed up in the survey participated in a 10-minutes interview through Zoom\footnote{\url{https://zoom.us/}} with the same questions as in the survey.

\begin{figure*}
\centering
  \includegraphics[width=\linewidth]{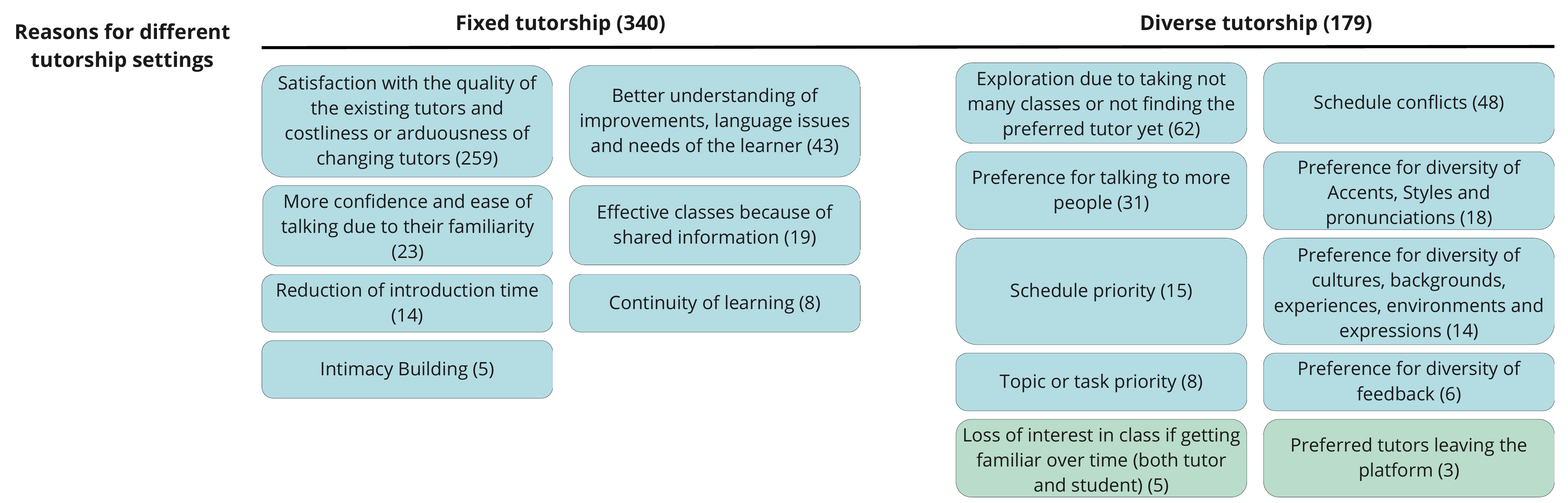}
  \caption{Reasons for different distributed tutorship settings from both the survey and interview results. 340 participants mentioned the reasons for having fixed tutorship, and 179 mentioned the reasons for having distributed tutorship. The reasons in the blue boxes come from the survey results, while the reasons in the green boxes are extra reasons added from the interviews. The numbers in the brackets represent the unique number of participants in the survey or the interview who mentioned those reasons.} 
  \label{fig:reasons}
\end{figure*}

Two of the authors extracted and coded all the responses from the survey and notes taken in the interviews. In total, we collected 677 reasons (some participants mentioned more than one reason) for choosing fixed or different tutors. A total of 16 responses are excluded because they are hard to understand. We further used the affinity diagram~\cite{hartson2012ux} to group the 677 reasons into 17 groups, as listed in Figure~\ref{fig:reasons}. The reasons in the blue boxes come from the survey results, while the reasons in the green boxes are extra reasons added from the interviews. The reasons for having fixed tutors are listed on the left, and the reasons for having different tutors are listed on the right.
The numbers in the brackets represent the unique number of participants in the survey and the interview who mentioned those reasons. 

\subsection{Results}

From Figure~\ref{fig:reasons}, we can see that most of the learners (340) preferred fixed tutorship compared to diverse tutorship (179). By looking at the top reasons for diverse tutorship, we can notice that some learners have not found preferred tutors or have schedule conflicts with their preferred tutors. These reasons might explain why the fixed tutorship has suggestive higher learning improvements in Section~\ref{sec:data}. The detailed reasons for each setting are described below. Learners on this platform are primarily adults who have the intermediate-to-advanced level of English.

The most popular reason to have fixed tutors is that they are satisfied with the current tutor, and it takes time and effort to change tutors and find suitable tutors. The second popular reason is that long-term tutorship allows tutors to understand learners' improvements, language habits/issues, and needs. For example, \textit{``The tutor whom I select regularly knows about my weaknesses and pronunciation habits. (P189)''} Furthermore, some learners thought familiarity with fixed tutors made them more confident and made it easier to speak during the sessions. In addition, some learners believed that with fixed tutors, they could have more effective classes based on the shared information in the previous sessions. For example, \textit{``through continuous classes with the tutor, background knowledge about each other is accumulated (P331).''} \textit{``Continuity of information sharing - conversation can be carried out efficiently (P429)''} In particular, a specific portion of learners thought having fixed tutors can save the introduction time in sessions. About eight learners mentioned that learning from fixed tutors can generally guarantee the continuity in learning, while five learners hoped to build intimacy with their fixed tutors.

The most popular reason to have various tutors is that learners are still in the exploration period due to having not taken many classes or found the preferred tutor yet. Many learners mentioned that they want to have the fixed tutorship but they have schedule conflicts with their preferred tutors. For example, \textit{``Even if you have a tutor you prefer, there's no class at the time you want (P369).''} In addition, many learners want to talk to more people and practice conversations with tutors with different accents or cultural backgrounds. Some learners care more about their schedule first, and they go for whoever is available at their preferred time. For example, \textit{``My schedule has priority over the tutor (P121).''} Others proposed that they focused more on the topics and want to find tutors who can discuss more on particular topics, for example, \textit{`` I made my selection according to conversation topic and purpose (P56).'' } A group of learners preferred the diversity of feedback from different tutors. Moreover, what is worth mentioning is that five learners during the interview said that both their and their tutors' interest had decreased in the class once they became too familiar with each other. For example, \textit{``If I meet one tutor continuously, then the tutor would become closer with me, and the variety of feedback decreases. Therefore, I change the tutor if I find another tutor who is nice (P47). ''} Lastly, a small portion of learners mentioned that their preferred tutors had left the platform because they worked part-time on the platform, they are thus learning from different tutors.



\section{Considerations of Learning under Distributed tutorship}

\begin{table*}[]
\small
\caption{Considerations for improving learners' learning under distributed tutorship from different parties.}
\label{tab:considerations}
\begin{tabular}{l|l}
\hline
                          & Solutions and Considerations                                                                                                                                                                                                                                                                                                                         \\ \hline
\multirow{3}{*}{Computer} & \begin{tabular}[c]{@{}l@{}}Summarize and quantify learners' learning history and share it with tutors:\\ 1. show aggregated data instead of detailed data because of privacy issues\\  2. show the data before the session instead of during the class in order not to distract the tutor\\  3. show top language issues of the learner\end{tabular} \\ \cline{2-2} 
                          & Recommend tutors that have similar personalities or domain knowledge to learners                                                                                                                                                                                                                                                                     \\ \cline{2-2} 
                          & Suggest scores or examples to tutors when they are giving the feedback                                                                                                                                                                                                                                                                               \\ \hline
\multirow{2}{*}{Tutor}    & Leave messages to subsequent tutors about what to pay attention to for a learner                                                                                                                                                                                                                                                                     \\ \cline{2-2} 
                          & Build profile and share information about a learner among tutors                                                                                                                                                                                                                                                                                 \\ \hline
\multirow{3}{*}{Platform} & Provide premium version to all learners book a certain number of recurring sessions                                                                                                                                                                                                                                                                  \\ \cline{2-2} 
                          & Support learners to build profiles that inform tutors about themselves                                                                                                                                                                                                                                                                               \\ \cline{2-2} 
                          & \begin{tabular}[c]{@{}l@{}}Coach tutors to minimize the quality variance \\ (e.g., giving high-quality feedback in a consistent format)\end{tabular}                                                                                                                                                                                                 \\ \cline{2-2} 
                          & \begin{tabular}[c]{@{}l@{}}Frame the tutor-computer collaboration carefully so that tutors do not feel technologies\\ will replace them\end{tabular}                                                                                                                                                                                                 \\ \hline
\end{tabular}
\end{table*}

To answer RQ3 (What are the implications for educators, tutors and platform designers to improve learners' language learning under distributed tutorship?), we conducted semi-structured interviews with three tutors and one product manager from Ringle.
\subsection{Semi-structured Interviews with Tutors and Product Manager}
We interviewed three tutors with an average of two years of English tutoring experience (E1: two years and three months, E2: three years and six months, E3: seven months) and the product manager (E4) from Ringle. Each interview lasted around 90 minutes, and the questions spanned three aspects: What are the reasons for the currently distributed tutorship phenomenon? What are the advantages and disadvantages of distributed tutorship for language learning? What are the possible solutions that can be applied to improve the learning experience on the online language tutoring platform?
We transcribed the interviews and extracted major themes by analyzing the interview text.

\subsection{Results}
\para{Reasons for Distributed Tutorship.} All three tutors and the product manager mentioned that online tutoring platforms share a similar business model to the gig economy, and distributed tutorship might be a byproduct of such a business model. In the gig economy, many people prefer to be independent contractors and freelancers for income and flexibility~\cite{abraham2018measuring}. As mentioned by E4, being undergraduates from top universities, it is hard for most tutors on Ringle to provide regular tutoring sessions, especially when semesters start or are close to the exam periods. E4 also said the platform tried to provide as many tutors as possible to satisfy the needs of language learners, and tutors on this platform are not required to have a fixed working time. E2 added that \textit{``modern professionals cannot provide a consistent schedule because they are multi-tasking.''} 

\para{Advantages and Disadvantages of Distributed Tutorship for Learners.}
The tutors pointed out both the advantages and disadvantages of distributed tutorship according to their experience. The advantage they mentioned most is the simulation of the real-world language learning environment: talking to different people. As E1 said, \textit{``For language learners, especially advanced learners, you have to talk about different things with different people. Language is amorphous.''} E3 added that different tutors have different strengths (e.g., writing skills or domain expertise), which can benefit learners in different aspects.
In terms of the disadvantages, they acknowledged that the new learning mode lacks continuity. E2 said that \textit{``the long-term relationship developed by the tutor and the tutee can help the tutee to overcome confidence issues and feel safe to try new things.``} Moreover, \textit{``when learning with the same tutor continuously, the tutor can take context into account when tutoring, e.g., the learners' performance drop is because of the lateness of the class at 11:00 pm; the pause is to make the speech more confident as told by the tutor last time instead of the lack of proficiency, and so on. 
And we can compare of the learners' current performance with their previous performance and know better about their habits and where to improve.''} However, E1 thought that though long-term tutorship might strengthen the personal tutor-tutee relationship, it may deteriorate the quality of feedback. 

\para{Considerations for Tutoring under Distributed Tutorship.}
We have summarized suggestions for future tutoring under ``Distributed Tutorship'' in three different categories based on the interview results:
computer techniques' perspective, tutors' perspective, and the platform's perspective. These are listed in the Table.~\ref{tab:considerations}. 

\textit{Computer.} 
Computer techniques can be utilized to quantify learners' learning history (e.g., learning activities and performance) and then share the data with tutors, so that they can give feedback and instructions accordingly to keep the learning continuity. In particular, E1 and E2 mentioned three points that need to be taken into consideration. First, the information shown can only be aggregated data instead of detailed conversation logs because of privacy issues. Second, it is better to display the data to the tutor before the class instead of during the session to avoid distraction. Third, to prevent information overload for tutors, the computer can extract the top language issues of the learner and share it with tutors. In addition,
computer techniques could also recommend tutors who have similar personalities and domain knowledge as the learners' preferred tutors. E4 thought that maintaining a pool of tutors with a good match is a more feasible solution than guaranteeing the same tutors from the platform's perspective.
Moreover, computer techniques can be applied to suggest scores or examples to tutors when giving the scores or feedback to guarantee that the feedback quality is consistent among different tutors (E1 and E4).

\textit{Tutor.} Tutors can leave useful information, i.e., messages for subsequent tutors and the learner's profile, to 
alleviate the learning discontinuity issue. For example, tutors can leave messages for subsequent tutors about aspects of the learner's personality, mentioned by E1, E2, and E3, and it may allow other tutors to be better prepared when running the session, by improving the efficiency of communication. In addition, tutors can build a profile for a learner and share the information with other tutors of this learner, which was mentioned by E2. For example, what knowledge or points need to be paid extra attention to when teaching the learner.

\textit{Platform.} The platform can provide services for more continuous learning experience. Firstly, the platform can provide services that help tutors and learners build a closer connection. The product manager (E4) suggested that the platform could provide premium products to allow learners to book a certain number of recurring sessions with the same tutor (e.g., 3-6 sessions), although
it is tough to provide permanent jobs for the tutors on the platform since many tutors are undergraduates who usually leave the platform after graduation. 
Furthermore, E3 proposed that the platform can support learners to build their profiles that inform tutors about themselves.
Secondly, the platform can provide services that minimize the quality variance among different tutors. For example, all (E1, E2, E3, and E4) stated that the platform could provide training for tutors to
give high-quality feedback in a consistent manner. 
Lastly, the platform should provide services encouraging the tutor-computer collaboration, instead of replacing tutors with computers. E2 emphasized that the platform and company should carefully frame the collaboration between tutors and computer techniques so that tutors do not feel technology will replace them. 


\section{Discussion}
In this section, we synthesize the findings from RQ1-RQ3, discuss the generalization and limitations of our findings, and share directions for future work to improve the learning experience under distributed tutorship.

\subsection{Learning and Teaching under Distributed Tutorship}
In this paper, we found that tutorship on the Ringle platform is highly distributed and the suggestive evidence that diverse tutorship might introduce lower learning improvement compared to fixed tutorship (RQ1). We also learned that learners found benefits in both fixed tutorship and diverse tutorship while learners highlighted the continuity and familiarity brought by fixed tutorship and the schedule conflicts with fixed tutors (RQ2). Together with the interviews with tutors and a product manager (RQ3), we suggest that techniques should be developed to increase consistency in the learning experience (e.g., tracking common errors across learning sessions) when the same tutor cannot be guaranteed. With learners exposed to more choices, learning techniques also need to be developed to support the matchmaking between tutors and learners. In addition, tools should also be designed to help tutors share the learners’ information with minimal privacy invasion. New challenges are brought to learning scientists and communities on how to analyze the learning experience involving both diversity and continuity in tutorship, and many other under-utilized factors, such as learning frequency and topics.

\subsection{Generalization}
Beyond language learning, other platforms and communities also have distributed tutorship dynamics and feedback culture, including P2P skill-sharing communities (e.g., Udacity\footnote{\url{https://www.udacity.com/}}, Clascity\footnote{\url{https://clascity.com/}}), and freelance markets (e.g., Upwork\footnote{\url{https://www.upwork.com/}}). These communities and platforms are developed to help individuals freely share their skills and receive feedback from one another. Users on these platforms also potentially experience distributed tutorship, such as learning to dance or write code from multiple tutors. 
Therefore, the considerations studied in this paper of how to unitize the benefits of diversity and flexibility while keeping the continuity of learning are far-reaching. For example, tutors can share and check the information of learners before starting a session.
Moreover, knowledge-sharing behaviors are becoming more widespread for the public. One example is people using TikTok\footnote{\url{https://www.tiktok.com/en/}}, a video social network service that has surpassed over 2 billion mobile downloads worldwide by October 2020, to share educational knowledge~\cite{fiallos2021tiktok, khlaif2021using}. Learning may become scattered and flexible in modern society, and connecting the dots and scattered learning fragments would be critical for learners. Our findings on improving the learning continuity can potentially enable more scattered and flexible learning options. For example, computer techniques can be utilized to keep track of learners' learning behaviors and share them with tutors or platforms to maintain the continuity in learning.

\subsection{Limitations and Future Work}
This work takes the first step toward applying learning analytics into the emerging learning mode of distributed tutorship, which we believe presents exciting opportunities for the learning analytics community.
Though we have found high-level patterns and suggestive evidence about the correlation between tutorship distributedness and the learning gains, a vast amount of work should be done further to investigate the details and more profound implications.
First, more analysis should be conducted to investigate learners' tutorship sequences to improve the reliability of the current results. In our current study, learners who have longer than 20 sessions are grouped into the mixed tutorship while they might have different stages in their tutorship sequences, e.g., exploration at the beginning and then fixed to some tutors, and then trying different tutors occasionally. We can apply moving windows on their learning history to recognize potential localized learning patterns. In addition, more detailed patterns can be investigated. For example, learners can have multiple fixed tutors when needed (e.g., $\langle\textit{aaabbbccc}\rangle$) or rotate (e.g., $\langle\textit{abcabcabc}\rangle$) instead of fixing on one tutor, and whether these strategies are beneficial for learning under distributed tutorship is worth investigating. 
Second, more learning data should be utilized to analyze the relationship with learning outcomes. For example, the time gap between learning sessions, the learning topics, the tutoring styles, and the content of tutors' feedback. Furthermore, we use tutors' scores to calculate the learning performance, which might be sensitive to tutors' individual scoring preferences. For example, in the fixed tutorship, the tutor might be more tentative to give a rising score as sessions accumulate.
Therefore, a computational method based on learning data (e.g., audio-to-text scripts) can be introduced to evaluate learners' language-speaking performance.
Third, more interviews or surveys can be conducted to understand the motivation behind learners' dynamic tutor selection behaviors across sessions. Fourth, distributed tutorship in other platforms and learning domains could be investigated. For example, how would the discontinuity of distributed tutorship in these settings affect learning? We appeal to researchers and educators from the education and computer science domains to pay attention to this new educational model as it is becoming increasingly popular and has many open questions for future research.

\section{Conclusion}
In this paper, we analyzed the emerging educational model of distributed tutorship within the context of an online language tutoring platform. Our data analysis shows that higher distributedness is suggestively correlated to lower learning gains. However, most learners have diverse tutorship or mixed tutorship instead of fixed tutorship on such a learning platform for various reasons (e.g., schedule conflicts). Our paper takes the first step to investigate distributed tutorship, and we hope our study could improve educators' and learners' awareness of the benefits and challenges of this emerging mode of learning. We believe that distributed tutorship will be a new trend for learning in the future due to its flexible and scalable nature. We also believe that the lack of continuity in tutorship needs to be further addressed, and the tutorship patterns we identified can guide future studies.

\begin{acks}
This work was supported by Institute of Information \& communications Technology Planning \& Evaluation (IITP) grant funded by the Korea government (MSIT) (No.2020-0-02237, Personalized Progress Analysis and Exercise Recommendation for Remote Language Learning Using AI and Big Data).
\end{acks}

\bibliographystyle{ACM-Reference-Format}
\bibliography{sample-base}


\end{document}